\documentclass[11pt]{article}

\usepackage{fullpage}

\newtheorem{theorem}{Theorem}
\newtheorem{corollary}{Corollary}%

\usepackage{graphicx}
\usepackage{algorithm}
\usepackage[noend]{algpseudocode}
\usepackage{ifthen}
\usepackage{color}

\newcommand{\dd }{\mathinner{.\,.}}
\newcommand{\no}[1]{}
\newcommand{\new}[1]{\textcolor{black}{#1}}

\newcommand{\access}{\ensuremath{\mathsf{access}}}
\newcommand{\rank}{\ensuremath{\mathsf{rank}}}
\newcommand{\select}{\ensuremath{\mathsf{select}}}
\newcommand{\writ}{\ensuremath{\mathsf{write}}}
\newcommand{\insertar}{\ensuremath{\mathsf{insert}}}
\newcommand{\delete}{\ensuremath{\mathsf{delete}}}
\newcommand{\suma}{\ensuremath{\mathsf{sum}}}
\newcommand{\search}{\ensuremath{\mathsf{search}}}
\newcommand{\leer}{\ensuremath{\mathsf{read}}}
\newcommand{\degree}{\ensuremath{\mathsf{degree}}}
\newcommand{\child}{\ensuremath{\mathsf{child}}}
\newcommand{\childrank}{\ensuremath{\mathsf{childrank}}}
\newcommand{\parent}{\ensuremath{\mathsf{parent}}}
\newcommand{\prev}{\ensuremath{\mathsf{prev}}}
\newcommand{\next}{\ensuremath{\mathsf{next}}}
\newcommand{\prefix}{\ensuremath{\mathsf{prefix}}}
\newcommand{\match}{\ensuremath{\mathsf{match}}}
\newcommand{\enclose}{\ensuremath{\mathsf{enclose}}}
\newcommand{\excess}{\ensuremath{\mathsf{excess}}}

\newcommand{\size}{\mathrm{size}}

\newcommand{\ones}{\mathrm{ones}}
\newcommand{\zeros}{\mathrm{zeros}}

\newcommand{\queries}{\mathrm{queries}}
\newcommand{\children}{\mathrm{child}}

\title{(Worst-Case) Optimal Adaptive Dynamic Bitvectors}

\author{Gonzalo Navarro\thanks{Department of Computer Science, University of
Chile}}

\date{}

\begin{document}

\maketitle

\begin{abstract}
While operations {\em rank} and {\em select} on static bitvectors can be
supported in constant time, lower bounds show that supporting updates raises
the cost per operation to $\Theta(\log n/ \log\log n)$ on bitvectors holding
$n$ bits. This is a shame in scenarios where updates are possible but uncommon. 
We develop a representation of bitvectors that we call adaptive dynamic
bitvector, which uses the asymptotically optimal $n+o(n)$ bits of space and, if 
there are $q$ queries per update, supports all the operations in 
$O(\log(n/q)/\log\log n)$ amortized time. Further, we prove that this time 
is \new{worst-case} optimal in the cell probe model.
We describe a large number of applications of our
representation to other compact dynamic data structures.
\end{abstract}

\section{Introduction}
\label{sec:introduction}

Bitvectors are the basic components of most compact data structures \cite{Nav16}.
Apart from the basic query $\access(B,i)$, which retrieves the bit $B[i]$ of 
the bitvector $B[1\dd n]$, they support two fundamental queries: 
\begin{description}
\item[$\rank_b(B,i)$,] which
tells the number of times the bit $b \in \{0,1\}$ occurs in $B[1\dd i]$, and
\item[$\select_b(B,j)$,] which gives the position of the $j$th occurrence of 
$b \in \{0,1\}$ in $B$.
\end{description}
 It is well known since the nineties that those 
operations can be supported in $O(1)$ time with a bitvector representation 
that uses $n+o(n)$ bits of space \cite{Cla96,Mun96}. 

Things are considerably
different, however, if we aim to allow updates to the bitvector: just 
supporting $\rank$ and bit flips requires $\Omega(\log n/\log\log n)$ time
\cite{FS89}. Indeed, one can support in $O(\log n/\log\log n)$ time,
and still within $n+o(n)$ bits, the operations \cite{NS14}
\begin{description}
\item[$\writ(B,i,v)$,] which sets $B[i]=v$, 
\item[$\insertar(B,i,v)$,] which inserts the bit value $v$ at position $i$ in
$B$, and 
\item[$\delete(B,i)$,] which removes the bit $B[i]$ from $B$.
\end{description}

This almost logarithmic gap between static and dynamic bitvectors permeates 
through most compact data structures that build on them, making dynamic
compact data structures considerably slower than their static counterparts, 
and not as competitive with classic data structures. Although this price
is in principle unavoidable, one may wonder whether it must be so high 
in cases where updates are sparse compared to queries, as
is the case in many applications. As an extreme example, since 
the static data structures can be built in linear time, one could have $O(1)$
amortized time if queries were $\Omega(n)$ times more frequent than updates,
by just rebuilding the static structure upon each update. The idea degrades
quickly, however: If queries are $q$ times more frequent than updates, this 
technique yields $O(n/q)$ amortized times.

In this paper we introduce a representation of dynamic bitvectors $B[1\dd n]$
that uses the asymptotically optimal $n+o(n)$ bits of space and offers 
$O(\log(n/q)/\log\log n)$ amortized time for all the operations, if queries 
are $q$ times more frequent than updates. We call our data structure {\em
adaptive dynamic bitvectors}.

To obtain our result, we modify classic dynamic bitvector representations 
\cite{RRR01,HSS03,CHL07,NS14}. Our structure is a tree of arity 
$\Theta(\sqrt{\log n})$
whose leaves may either be ``dynamic'', storing $\Theta(\log^2 n/\log\log n)$ 
bits and supporting updates, or long ``static'' bitvectors handling only 
queries. A whole subtree is converted into static---which we call
``flattening''---when it has received sufficient queries to amortize the
cost of building the static structures (i.e., linear in the number of bits it 
represents). When an update falls in a static leaf, the leaf is recursively
split into static leaves of decreasing lengths along a path towards the
position to modify, until a (short) dynamic leaf is produced and the update
is executed there---a process we call ``splitting''. For maintaining balance 
in the tree we build on Weight-Balanced B-trees (WBB-trees) \cite{Die89,AV96}, 
which interact well with our new operations of flattening and splitting. 

We also prove that the \new{(amortized)} time complexity of our data structure 
is optimal \new{in the worst case}. For
this sake, we modify Fredman and Saks' \cite{FS89} $\Omega(\log n/\log\log n)$
lower bound, on the cell probe model, for the time of a sequence of operations
$\writ$ and $\rank$---actually a simpler version that queries only the parity
of the rank---, to consider the case where the frequency of updates is $1/q$.
We obtain the lower bound $\Omega(\log(n/q)/\log\log n)$.

Finally, we describe how our adaptive dynamic bitvectors can be used to
improve the time complexity of a number of dynamic compact data structures that build
on bitvectors, like arrays, sequences, trees, texts, grids, graphs, and others.

A preliminary partial version of this paper appeared in {\em SPIRE 2024}
\cite{Nav24}. In this extended version we significantly improve the space and
time of our data structure, prove its \new{(worst-case)} optimality, and 
explore its applications in depth.

\section{Our Work in Context} \label{sec:related}

Our problem is an instance of the so-called ``dynamic bitvector with indels''
problem, which as said requires $\Omega(\log n/\log\log n)$ time per operation
even if we support only $\rank$ and $\writ$ \cite{FS89}. Several solutions have
matched 
this lower bound, or been close to. Hon et al.~\cite{HSS03} store a dynamic 
bitvector $B[1\dd n]$ in $n+o(n)$ bits of space, handling queries in time 
$O(\log_b n)$ and updates in time $O(b)$, for any $b = 
\Omega((\log n/\log\log n)^2)$. Their main structure is a WBB-tree
\cite{Die89,AV96}. Chan et al.~\cite{CHL04} use balanced binary trees 
with leaves containing $\Theta(\log n)$ bits, 
obtaining $O(n)$ bits of space and $O(\log n)$ time for all the operations. 
M\"akinen and Navarro \cite{MN08} still use balanced binary trees, but use
leaves of $\Theta(\log^2 n)$ bits, retaining their $O(\log n)$ times but 
reducing the space to $n+o(n)$ bits. Finally, Navarro and Sadakane \cite{NS14} 
replace binary trees by structures closer to B-trees, retaining the $n+o(n)$ 
bits of space and supporting all the operations in the optimal time 
$O(\log n / \log\log n)$. In those terms the problem is regarded as closed.

In this paper we focus on a regime, however, that is relevant for many
applications: we assume that there are, on average, $q$ queries per update.
In this regime, we obtain $O(\log(n/q)/\log\log n)$ time for all the
operations, which we prove to be optimal \new{in the worst case. Our time} is 
amortized, as we rely on converting whole subtrees into static structures 
(which answer queries in $O(1)$ time) when they have received sufficient 
queries to pay for that conversion. The conversion needs to temporarily copy 
the bits stored in the converted subtree, but we still manage to use $n+o(n)$
bits of space.

There has been work to store the bitvectors within entropy space, which
means $Hn$ bits with $H =
\frac{m}{n}\log_2\frac{n}{m}+\frac{n-m}{n}\log_2\frac{n}{n-m}$, $m$ being the 
number of $1$s in the bitvector. Assuming $m < n/2$, Blandford and 
Blelloch \cite{BB04} obtain $O(nH+\log n)$ bits of space while supporting all 
operations in $O(\log n)$ time, using a balanced binary tree where the distances
between consecutive 1s are gap-encoded in the leaves. M\"akinen and Navarro
\cite{MN08} improve the space to $nH + o(n)$ bits, while retaining $O(\log n)$
time for the operations. Navarro and Sadakane \cite{NS14} retain this space
and reduce the time to the optimal $O(\log n/\log\log n)$. We discuss in
Section~\ref{sec:app} how our results can be extended to use entropy-bounded 
space.

\section{Weight-Balanced B-trees (WBB-trees)} \label{sec:wbb}

We briefly survey in this section the classic data structure ours builds on. 
A {\em Weight-Balanced B-tree (WBB-tree)} \cite{Die89,AV96} is, like a B-tree, a
multiary balanced tree where all the leaves are at the same depth. Given some 
node arity $a \ge 16$ and leaf size $b$ multiple of 16, the WBB-tree guarantees
that every internal node
has $a/4$ to $4a$ children (save the root, which has $2$ to $4a$ children) and
leaves store $b/4$ to $b$ elements.

As said, all the leaves of the WBB-tree are at the same level, which is called 
level $0$. The parent of a level-$l$ node is of level $l+1$. We call
$\size(v)$ the number of elements in leaves descending from node $v$. 
The key WBB-tree invariant is the so-called {\em weight-balancing constraint}:
every node $v$ at level $l$, except possibly the root, satisfies
$a^l b/4 \le \size(v) \le a^l b$. This implies that the height of a 
WBB-tree storing $n$ elements is 
$h \le 1+\log_{a/4} (n/(b/4)) = \Theta(\log_a(n/b))$.

The navigation of WBB-trees is similar to that of B-trees: internal nodes
store up to $4a$ routing keys, which are used to search for a key from the
root towards the proper leaf. The total number of nodes visited by a search or
update operation
is then $h+1$. As seen later in the paper, it is convenient for now to ignore
the time spent by searches at each internal node or leaf.

The WBB-tree enforces the weight-balancing constraint and the bounds on leaf 
sizes; the
arity bounds on internal nodes come as a consequence. The insertion at a leaf
may make it overflow, that is, store more than $b$ elements. The leaf is then
cut into two leaves of similar sizes, which are always between $\frac{7}{16}b$
and $\frac{10}{16}b$. While this increases the arity of the parent, the
WBB-tree does not directly control internal node arities, as said. It only 
enforces the weight-balancing constraint. When an insertion makes an internal
node $v$ of level $l$ have $\size(v) > a^l b$, it is cut into two siblings,
balancing their sizes as much as possible. It can be seen that both sizes are
between $\frac{6}{16} a^l b$ and $\frac{11}{16} a^l b$.

Deletions can produce leaf underflows, that is, they may be left with less
than $b/4$ elements. In this case they are merged with a sibling leaf and then,
if necessary, cut again into two of about the same size. The resulting leaf
sizes are between $\frac{8}{16}b-1$ and $\frac{14}{16}b-1$. Similarly, when
the size of an internal node at level $l$ falls below $a^l b/4$, it is merged
with a sibling node in the same way, so that the resulting sizes are between
$\frac{8}{16}a^l b-1$ and $\frac{14}{16}a^l b-1$.

In all cases, the cost of correcting a violation of the weight-balancing
constraint is $O(a)$ for internal nodes and $O(b)$ for leaves. An insertion or
deletion visits $h$ internal nodes and a leaf, and because of the corrections
it may cost $O(a h + b)$ time in the worst case. This may occur because, 
when we insert or delete an element in a leaf, we may have to correct all
the nodes in the return path, if their (leaf or internal node) sizes fall out
of bounds. An important property of WBB-trees is that, once a correction takes 
place at level $l$, the node needs to receive $\Omega(a^l b)$ further updates 
in order to need a new correction. In an amortized sense, then, the $O(b)$ and
$O(a)$ cost of maintaining leaves and internal nodes, respectively, within the 
allowed sizes can be absorbed by charging just $O(1)$ time to the visit of the
update operations to each node. This makes the {\em amortized} cost of updates 
just $O(h + b)$ (we always pay $O(b)$ time at leaves to insert the new element).

\section{Adaptive Dynamic Bitvectors} \label{sec:adb}

We use the transdichotomous RAM model of computation, with computer words of 
$w$ bits, so we can handle in memory bitvectors of length up to $2^w$.
We call $n \le 2^w$ the current length of the bitvector $B$. We do not use
$w$-bit systemwide pointers, but pointers of $O(\log n)$ bits. This requires
specialized (though not very complex) memory management techniques
\cite{Mun86} \cite[Sec.~4.6]{MN08}; in particular allocation and deallocation 
takes constant time). It also requires
us to assume that, say, $\lceil \log_2 n\rceil$ stays constant. We handle this
by completely rebuilding the data structure when $\lceil \log_2 n\rceil$
increases by one or decreases by two, which adds only $O(1)$ time to the
amortized cost of updates. 

Our structure uses a multiary tree, much as in previous work that obtained
$O(\log n/\log\log n)$ time for all the bitvector operations
\cite{RRR01,HSS03,CHL07,NS14}.
The main novelty is that, in order to speed up queries when updates are
scarce, we convert some subtrees to static leaves that handle queries in
constant time. Static leaves at high levels make queries to their positions
faster, as those queries traverse a short path and end in constant time on a
static leaf. We have chosen WBB-trees because they allow us obtain the desired
amortized times.

\subsection{Structure}

As anticipated, our data structure is a modified WBB-tree with parameters
$a = \max(16,\lceil \sqrt{\log_2 n}\rceil)$ and $b = 16 \lceil \log_2^2
n / (16 \log_2\log_2 n)\rceil$. We are not using the WBB-tree to store
elements that are searched for by value, but to store bits, on which we want to
support the queries $\access$, $\rank$, and $\select$, and the updates 
$\writ$, $\insertar$ and $\delete$. Our tree leaves are of two types:

\begin{itemize}
\item A ``dynamic leaf'', which corresponds to the WBB-tree leaves. It stores
$b/4$ to $b$ bits and no $\rank$/$\select$ precomputed answers. A dynamic leaf 
answers $\access$ queries in $O(1)$ time and $\rank/\select$ queries in time
$O(b/\log n) = O(\log n/\log\log n)$, via sequential scanning (details are 
given soon).
\item A ``static leaf'', which can appear at any level $l>0$ and stores 
arbitrarily large bitvectors, with their corresponding precomputation to solve 
$\access/\rank/\select$ queries in $O(1)$ time \cite{Cla96,Mun96}.
\end{itemize}

The internal tree nodes $v$ record the following fields (each entry in
$O(\log n)$ bits):
\begin{description}
\item[{$v.\children[1\dd 4a]:$}] the up to $4a$ children of $v$.
\item[{$v.\size[1\dd 4a]:$}] the numbers $v.\size[i] = \size(v.\children[i])$ of
bits below each child of $v$.
\item[{$v.\ones[1\dd 4a]:$}] the number of 1-bits below each child of $v$.
\item[{$v.\zeros[1\dd 4a]:$}] the number of 0-bits below each child of $v$.
\item[$v.\queries:$] number of queries ($\access$/$\rank$/$\select$) that 
traversed $v$ since the last update ($\writ$/$\insertar$/$\delete$)
that traversed $v$, or since the creation of $v$.
\end{description}

The arrays $v.\size$, $v.\ones$, and $v.\zeros$ in each internal node $v$
will be maintained using a data structure of Raman et al.~\cite{RRR01}. For
any sequence $X = x_1,x_2,\ldots,x_{4a}$ of length $4a= O(\sqrt{\log n})$, 
of $O(\log n)$-bit numbers, the structure uses $O(a\log n)$ bits of space and
is built in $O(a)$ time (we use zeros for the positions of $X$ that are beyond 
the current node arity). The structure supports increasing or decreasing any 
$x_k$ by up to $O(\log n)$ in $O(1)$ time, and 
computes also in $O(1)$ time $\suma(X,k) = \sum_{t=1}^k x_t$ for any $k$ and
$\search(X,s) = \min \{ k, \suma(X,k) \ge s\}$ for any $s$.

\subsection{Queries}

The queries use in principle the standard mechanism for dynamic bitvectors
$B$ \cite{RRR01,HSS03,CHL07,NS14}. For $\access(B,i)$, we descend from the WBB-tree root
to a leaf. At each node $v$, we compute the index $k=\search(v.\size,i)$ of
the child to descend by. We then descend to $v.\children[k]$, updating $i
\gets i - \suma(v.\size,k-1)$. Upon reaching a leaf, we read its $i$th bit in
constant time. The worst-case time is then $O(h)=O(\log_a (n/b))
= O(\log n / \log\log n)$.

For $\rank_1(B,i)$ we proceed analogously, except that we also start with a
counter $r \gets 0$. Each time we descend to $v.\children[k]$, we increase $r
\gets r + \suma(v.\ones,k-1)$. At the leaf, however, we must count the $o$ 1s 
up to position $i$, so as to return $r+o$. To compute $\rank_0(B,i)$ we just
return $i-\rank_1(B,i)$.

The leaf can be scanned in time $O(b/\log n)$, by processing chunks of
$\Theta(\log n)$ bits in constant time. This is done via small 
tables of $O(\sqrt{n})$ entries, which precompute the number of 1s in every
possible chunk of $\frac{1}{2}\lceil \log_2 n\rceil$ bits. The tables are
rebuilt in $o(n)$ time whenever we reconstruct the whole structure because
of changes in $\lceil \log_2 n\rceil$, so their overhead is negligible.
We then count the 1s of consecutive chunks of $\Theta(\log n)$ bits in the 
leaf, until we reach the chunk that contains position
$i$. To count the 1s in a prefix of that chunk in constant time, we use a
slightly larger table of size $O(\sqrt{n}\log n)$ that counts the 1s in every
prefix of every chunk; this still takes $o(n)$ bits of space and construction
time. 

The worst-case time of $\rank$ is then $O(h+b/\log n) = O(\log n / \log\log n)$.

The solution to $\select_1(B,j)$ is the dual of that of $\rank_1(B,i)$. We
start with a counter $p \gets 0$ and compute $k = \search(v.\ones,j)$ to 
find the child to descend by. We then set $j \gets j - \suma(v.\ones,k-1)$ and 
$p \gets p + \suma(v.\size,k-1)$. For $\select_0(B,j)$ we proceed identically,
using $v.\zeros$ instead of $v.\ones$.

At the static leaf, we must still find the $j$th $1$ or $0$. We scan the leaf
chunk-wise with the same table used for $\rank$, until we identify the chunk
that contains the answer. We then make use of new tables of $O(\sqrt{n}\log
n)$ entries that give the position of the $j$th $1$ and the $j$th $0$ inside
every possible chunk, for every $j$. The worst-case time of both $\select$
operations is then also $O(h+b/\log n) = O(\log n / \log\log n)$.

In our data structure, however, the three queries may end up at a static leaf.
Static leaves have precomputed the data structures
\cite{Cla96,Mun96} that solve $\rank$ and $\select$ in constant time.
In those cases, the time of all the queries is $O(h-l)$, where $l$ is the
level of the static leaf arrived at. For example, as a glimpse of our final
result, if the static leaf is of size $\Theta(q)$, then by the WBB-tree
invariants it is at level $l = \log_a(q/b)+O(1)$, and the query times are
$O(h-l) = O(\log_a(n/q)) = O(\log(n/q)/\log\log n)$.

\paragraph{Flattening}

The novelty in our adaptive scheme is that, every time we traverse an internal
node $v$ for any of the three queries, we increment $v.\queries$, and if we
traverse it for an update, we reset $v.\queries$ to zero. If, after receiving
a query, it holds $v.\queries \ge v.\size$, we convert the whole subtree of 
$v$ into a static leaf, which we call
``flattening'' $v$. Flattening is done in time $O(v.\size)$, by traversing and 
deleting the subtree of $v$, while writing the bits of all the leaves onto a 
new bitvector, which is finally preprocessed for constant-time queries and 
converted into the static leaf corresponding to $v$.
We show later, however, that its amortized 
cost is absorbed by the preceding $v.\size$ queries.

Note that flattening temporarily increases the space usage by $v.\size$ bits, 
which may be as much as $n$ if $v$ is the root (we reduce this impact later). 
Note also that flattening does 
not change $v.\size$, and thus it does not affect the weight-balancing
constraint of the WBB-tree.

\subsection{Updates}

Updates are handled, in principle, as in previous work \cite{RRR01,HSS03,CHL07,NS14}. To
perform $\writ(B,i)$, we traverse the tree as for $\access$, 
modify the corresponding bit in the (dynamic) leaf we arrive at (we consider 
soon the case where we arrive at a static leaf), and increase or decrease
at most one entry of $v.\ones$ and $v.\zeros$ for each node $v$ as we
return from the recursion. Because those updates to $v$ take constant time
\cite{RRR01}, the operation takes time $O(h) = O(\log n/\log\log n)$.
Note that $\writ$ has no effect on the tree balance. Instead, 
it does reset $v.\queries$ to zero on the traversed nodes $v$.

Insertions and deletions are analogous, yet at the end they insert or delete
a bit in a dynamic leaf and must also update $v.\size$ along the path. 
The bits of a leaf can be shifted in time $O(\log n / \log\log n)$, by chunks of
$\Theta(\log n)$ bits (by using tables or word-wise RAM operations,
which shift and copy chunks of $w \ge \log_2 n$ bits in constant time).
The insertion and deletion of bits at leaves may cause violations on the
allowed leaf and internal node sizes. We have seen in Section~\ref{sec:wbb}
that those can be handled by charging a constant to the amortized time spent
by the update operations on each internal node, so we do not need to consider
them further. The total amortized time for updates is then $O(h+b/\log n) =
O(\log n/\log\log n)$.

\paragraph{Splitting}

\begin{figure}[t]
\includegraphics[width=\textwidth]{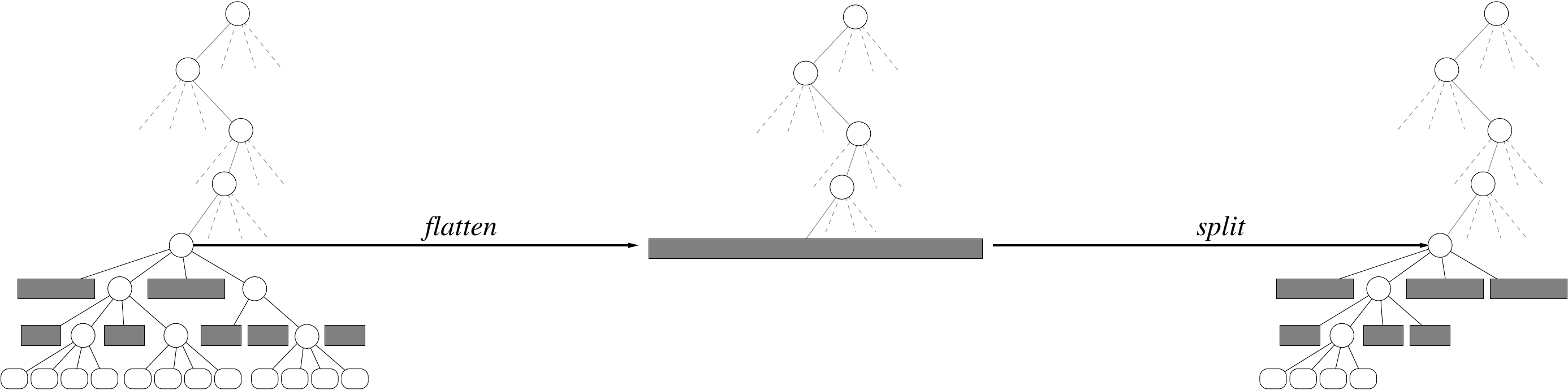}
\caption{Flattening and posterior splitting of a node, the former after
receiving over $v.\size$ consecutive queries, and the latter 
when receiving an update at a static leaf; the leaf is recursively split
until the update falls in a dynamic leaf. Circles are
internal nodes, round rectangles are dynamic leaves, and gray rectangles are
static leaves.}
\label{fig:splitflat}
\end{figure}

The novel part of updates occurs when we arrive at a static leaf $v$. In
this case, we ``split'' $v$, which means replacing it by an internal node of 
arity $a$. From its children, $a-1$ are static leaves and the one containing 
the position to update is internal.
Note that, if $v$ is of level $l$, then it holds $a^l b/4 \le v.\size \le 
a^l b$, and therefore the children, which are of level $l-1$, have sizes
$\lfloor v.\size/a \rfloor$ and $\lceil v.\size/a \rceil$, which are between
$a^{l-1}b/4$ and $a^{l-1}b$. Thus, the children maintain the 
weight-balancing constraint. The internal node created at level $l-1$ is, in
turn, split into $a$ children in the same way, until re reach the internal
node of level $1$ where the update must be applied. This node, which represents
$ab/4 \le t \le ab$ bits, is made of arity $a' = (3/4)t/b$, so that the 
leaves are created with size $(3/4)b$; note $a/3 \le a' \le (4/3)a$ is a valid 
arity. We can now finally perform the update on the proper leaf.
Figure~\ref{fig:splitflat} illustrates flattening and splitting.

Splitting $v$ takes $O(v.\size)$ time. The complete sequence of splits up to
the leaves also takes total time $O(v.\size)$, as we create consecutive nodes
of exponentially decreasing lengths, at most $a^{l'} b$ per level $l' = l,
l-1, \ldots, 0$. Though the worst-case cost of updates can then be $\Theta(n)$,
we prove sublogarithmic amortized bounds later.

An issue that arises when combining the WBB-tree corrections of
Section~\ref{sec:wbb} with our static leaves is
that, while the weight-balancing constraint can only be violated on an
internal node (static leaves always satisfy the constraint), the merging
procedure might need to merge a level-$l$ internal node whose size fell below
$a^l b/4$ with a sibling node $u$, and that node $u$ may be a static leaf.
In this case, we split $u$ into $a$ children, which are all static
leaves if $l>1$ and dynamic leaves otherwise (in which case we might create
$a'$ leaves, as above). Then the merging is carried out. Since, as shown in
Section~\ref{sec:wbb}, these mergings occur every $\Theta(a^l b)$ updates, the 
$O(a^l b)$ cost of this splitting is still absorbed by the $O(1)$ cost we
already charge to every update that visits $v$.

\paragraph{Reducing space}

In order to maintain the total space in $n+o(n)$ bits (plus the space for
flattening), we store the bits at the leaves $v$ using exactly $\lceil
v.\size / \log_2 n \rceil$ chunks of $\lceil \log_2 n\rceil$ bits. 
We reallocate the leaves as needed when bits are inserted or deleted in order 
to maintain this invariant. The cost of reallocation is already included in 
the $O(b/\log n)$ cost of bit insertion or
deletion. The invariant ensures that we waste only $O(\log n)$ bits per leaf, 
which amounts to $O(n\log\log/\log n)=o(n)$ total bits, given that the
WBB-tree invariants ensure that there are $O(n/b)$ leaves overall.

To ensure the desired maximum space of $n+o(n)$ bits, we avoid flattening nodes
$v$ where $v.\size > n/\lceil \log_2 n\rceil$, so the maximum temporary space
for flattening is $O(n/\log n)$ bits and the space is within $n+o(n)$ bits. 
We show in the next section that this does not affect the time complexities.

\section{Amortized Analysis} \label{sec:anal}

We will use an accounting scheme to prove that the amortized cost of all the
adaptive dynamic bitvector operations described in Section~\ref{sec:adb}
is $O(\log(n/q)/\log\log n)$. 

We will use a model where a {\em node} will refer to a particular WBB-tree node
in the lifetime of the data structure, from the moment in which it is created
until the time it is destroyed. Flattening or splitting a node counts as
destroying it and creating a new node of another type. Further, every time an 
update traverses an internal node $v$ and resets its counter $v.\queries$ to 
zero, this counts in our model as destroying the node and creating a new one.
So our nodes are created, receive queries, and then disappear.

The level of a node does not change along its lifetime. We define ${\cal I}_l$
as the set of all internal nodes of level $l$ that have existed along the
lifespan of the data structure. We assume that $m$ queries and $m/q$ updates 
occur along this lifespan (we should actually use 
$m/(q-1)$ to have an average update frequency of $1/q$, but this way yields
cleaner formulas; the big-O results are of course the same).

We note that queries create only static leaves, by flattening. Only updates 
can create internal nodes, by zeroing $v.\queries$ (according to our model of
node in this section), by correcting internal nodes, or by splitting static 
leaves.
Note that an update can create at most two internal nodes per level (this
happens when cutting nodes, otherwise they
create just one). Each update starts at the root and ends at a dynamic leaf. 
Since there are $m/q$ updates, the nodes of each level $l$ are visited by $m/q$ 
updates, and so at most $2m/q$ internal nodes can be created in level $l$ along 
the lifespan of the data structure, that is, it holds
$|{\cal I}_l| \le 2m/q$ for every $l$ if the tree starts empty.

\paragraph{Queries and flattening}

Queries cost $O(1)$ time per internal node traversed. We charge that cost
to those internal nodes. In case the query arrives at a static leaf, it spends
$O(1)$ further time on the leaf, which is charged to the query. If the query
arrives, instead, at a dynamic leaf, it spends $O(\log n/\log\log n)$ time on 
it. Dynamic leaves, however, are of level $0$; therefore we can distribute the 
$O(\log n/\log\log n)$ cost spent at the dynamic leaf over the $h = \Theta(\log
n/\log\log n)$ internal nodes traversed to reach it. This increases the 
cost charged to internal nodes traversed by just another $O(1)$.

The cost of flattening is also charged to the flattened node. An internal node
$v$ is flattened once $v.\size$ consecutive queries have traversed
it, and flattening costs $O(v.\size)$. Therefore, we charge $O(1)$
additional cost to the nodes traversed by queries to pay for their eventual
flattening.

So far, we have charged $O(1)$ to query operations and zero to flattening,
and charged most of the actual cost to internal nodes, $O(1)$ per time a query
traverses them. We now calculate how 
much can be charged to all the internal nodes ${\cal I}_l$. 

An internal node $v$ can be traversed by $v.\size$ queries before it is
flattened (and thus destroyed; recall that in our model updates destroy the 
nodes they visit and create new versions). If $v \in {\cal I}_l$, then 
$v.\size \le a^l b$; 
hence it can be charged by queries at most $a^l b$ times before it gets 
flattened.

Since there are $m$ queries, and each visits each level at most once, there
are at most $m$ queries affecting the nodes of each ${\cal I}_l$. On the other 
hand, $|{\cal I}_l| \le 2m/q$. Since those nodes can be charged at most 
$a^l b$ times before disappearing, we could distribute all the
charges of the $m$ queries only if $m \le (2m/q) a^l b$, that is, if 
$l \ge l^*=\log_a(q/2b)$. Across those levels $l$, the total charges
to nodes add up to $m (h-l^*) = O(m\log(n/q)/\log\log n)$.

The intuition is that, in the higher levels ($l \ge l^*$), we can distribute 
$q/2$ queries to each of the $2m/q$ nodes; they are large enough to receive 
those $q/2$ 
queries before flattening. On the deeper levels, instead, it is not possible 
to assign all the $m$ charges to nodes before they flatten, which means that 
not all queries can reach those deep nodes, because they inevitably flatten 
the nodes at level $l^*$. Precisely, the nodes at level 
$l^*$ are of length $\le q/2$ and thus might receive $q/2$ queries, but 
from there on, the $2m/q$ nodes at levels $l < l^*$ can receive at most 
$(q/2)/a^{l^*-l}$ queries before flattening. Adding up over all levels 
$l < l^*$, we obtain that the amount of queries that can be received is at
most $2m/q \cdot O(q) = O(m)$.

Overall, the $m$ queries, including the induced flattenings, have an amortized
cost of $O(\log(n/q)/\log\log n)$ per operation.

\paragraph{Updates and splitting}

Updates do not finish at static leaves; they open up a path to a dynamic
leaf is necessary. They then take amortized time $O(h+b/\log n)=
O(\log n/\log\log n)$. Since there are $m/q$ updates, however, their total
contribution is just $O((m/q)\log n/\log\log n) \subseteq
O(m\log(n/q)/\log\log n)$, or $O(\log(n/q)/\log\log n)$ per operation. 

An update can also, however, split a flattened leaf, creating a path of 
internal nodes and static leaves, until reaching level $0$. We have already 
accounted for the internal nodes splittings create; let us now analyze the 
cost of those splittings. Note that we do not refer to the splittings induced
by WBB-tree corrections; the cost of those are already accounted for within
the formula of the preceding paragraph. Each update operation can produce one
split per level.

Let us call ${\cal S}_l$ the set of static leaves of level $l$ that have
existed along the lifespan of the data structure. Let us assume
pessimistically that $|{\cal S}_{l^*}| \ge m/q$. Then, in our quest to maximize
the cost of splittings, we will never choose to flatten a leaf of level below 
$l^*$, because we have sufficiently many of them at level $l^*$ to choose 
from and those are costlier to split. 
Leaves of level $l^*$ are of length at most $q/2$, therefore, if we use 
our at most $m/q$ splittings on those, the total splitting cost would be at 
most $m/2$.

This cost is not maximal, because we are splitting only flattened leaves, whose
total length is indeed bounded by $m$. To achieve higher splitting costs, we 
must split some flattened leaves and then the static leaves created by the 
splitting, several times.

Let us then choose a set ${\cal F} = \{ v_1,v_2,\ldots\}$ of static leaves 
created by flattening by the $m$ queries that occurred along the lifespan of
the data structure. Let $\ell_i$ be the length of leaf $v_i$, so
$\sum_i \ell_i \le m$. Each split on some $v_i$ creates a number of
new static leaves (not in $\cal F$), on which other splits may apply later.

To bound how much can we pay by splitting the leaves created from some flattened
leaf $v \in \cal F$ at level $l$, assume the maximum size $v.\size = \ell = 
a^l b$. The first split costs $\ell$, and creates $a-1$ static leaves of 
lengths $\ell/a$, $a-1$ of length $\ell/a^2$, etc. The next $a-1$ splits
pessimistically choose leaves of size $\ell/a$, costing $\ell/a$ and creating
$a-1$ leaves of size $\ell/a^2$, $a-1$ of size $\ell/a^3$, etc. Now there are
$a(a-1)$ leaves of size $\ell/a^2$ ($a-1$ created with the first split and
$(a-1)(a-1)$ created during the next $a-1$ splits), which are the next ones to
choose to maximize costs, and so on. Let $L(t)$ be the number of leaves of
length $\ell/a^t$ created in the process. It can be seen that
$L(t) = a^{t-1}(a-1)$. The sum of all the leaf lengths for any $t>0$ is then
at most $\ell (a-1)/a$. Therefore, if we split all the static leaves from $v$
along $t$ levels, the total splitting cost is $1+(\ell(a-1)/a) t \le \ell t$.

By our previous observation, we never choose to split nodes beyond level
$l^*$, so $t \le l-l^* \le h-l^*$.
The total splitting cost is then at most $\sum_i \ell_i (h-l^*)
\le m(h-l^*) = O(m\log_a(n/q)) = O(m\log(n/q)/\log\log n)$.

The amortized cost of all our operations is then $O(\log(n/q)/\log\log n)$. 
Finally, to use only $O(n/\log n)$ temporary bits of space for flattening,
we do not flatten nodes $v$ where $v.\size > n/\lceil \log_2 n\rceil$. The
level of those nodes is at least $l_0 = \log_a(n/(b\lceil \log_2 n \rceil))$.
The number of (highest) levels that are not flattened is then $h-l_0 = 
O(\log_a(\log n)) = O(1)$. This changes the costs only by an additive constant.

Finally, the $o(n)$ term in the space is $O(n\log\log n/\log n)$ due to the 
space
wasted in leaves and internal nodes. The sublinear term in the static data
structures can also be made $O(n\log\log n/\log n)$ while retaining linear
construction time \cite{Gol06,RRR07}.

\medskip
\begin{theorem} \label{thm:1}
An adaptive dynamic bitvector starting empty can be maintained in
$n+O(n\log\log n / \log n)$ bits of space, 
where $n$ is the current number of bits it represents,
so that if the fraction of updates over total operations so far is $1/q$,
then the bitvector operations take $O(\log(n/q)/\log\log n)$ amortized time.
\end{theorem}
\medskip

If starting on a tree of $n_0$ nodes, we must add $O(n_0\log n_0/\log\log n_0)$
to the cost of the whole sequence of operations, so as to simulate the first 
$n_0$ insertions; the analysis holds if $m/q = \Omega(n_0/b)$ (actually, it
suffices that $\log(m/n_0) = \Omega(\log(q/b))$.


\section{A Matching Lower Bound} \label{sec:lbound}

In this section we prove that our algorithm is indeed optimal, by slightly 
adapting the $\Omega(\log n / \log\log n)$ proof of Fredman and Saks
\cite{FS89}. We actually follow an excellent and unpublished survey of
Miltersen \cite{Mil00}, which gives a much cleaner proof. As the material is
unpublished and not available at a formal repository, we repeat its details in 
what follows, modifying the proof to consider the assumption that the fraction 
of updates is $1/q$, applying minor fixes and improvements, and explaining it in 
more depth.

Assume the RAM word size is $w=\Theta(\log n)$ and that the updates are only of the
form $\writ(B,i)$. Further, consider a simpler variant of the problem where
the queries are of the form $\prefix(B,i) = \rank(B,i) \bmod 2$ (this is called
the dynamic prefix problem, which obviously reduces to the dynamic bitvector 
problem). 

Assume $n$ is a power of $2$ and
consider $k=n/q$ updates\footnote{This number is crucial to
obtain the result. It is $\sqrt{n}$ in Miltersen's proof, but we do not have
$\sqrt{n}$ updates if $q > \sqrt{n}$, that is, if there is a chance that the
complexities $\log n$ and $\log(n/q)$ differ.} $\writ(B,i_k,a_k),$
$\writ(B,i_{k-1},a_{k-1}), \ldots,$ $\writ(B,i_1,a_1)$, 
where the positions are distributed
in rounds as $i_1 = n/2$ (round 1), then $i_2 = n/4$ and $i_3 = 3n/4$ (round 2),
then $i_4=n/8$, $i_5=3n/8$, $i_6=5n/8$, and $i_7=7n/8$ (round 3), and so on. 
It is easy to see that, if $u,v \le r$ for some $r$, then two different write 
positions $i_u$ and $i_v$ are sufficiently distant, that is,
$|i_u-i_v| \ge n/(2r)$ (and also $n-i_u \ge n/(2r)$).

The sequence of writes is divided into ``epochs''. Epoch 1 is formed by the
last $l_1=\log^3 n$ $\writ$ operations (recall that we apply the writes in
reverse order, so epoch 1 contains the writes at positions $i_{l_1}, i_{l_1-1},
\ldots, i_1$). Epochs 1 and 2 contain the last $l_2=\log^6 n$ $\writ$ 
operations, that is, epoch 2 spans the writes at positions $i_{l_2}$ to 
$i_{l_1+1}$). In general, epochs 1 to $i$ contain the last $l_i=\log^{3i} n$
$\writ$ operations. We call $r$ the number of epochs, which satisfies 
$\log^{3r} n = k$, that is, $r = \log (n/q) / (3\log\log n)$ \new{(we can 
assume $q \le n/\log^3 n$ so that $r \ge 1$; for larger $q$ the lower bound we 
derive is already $1$)}.

The cells in memory will be ``stamped'' with the last epoch where they were
written. Epochs will define the granularity of our analysis; we will show that 
many queries need to read one cell from \new{each} epoch; the lower bound then
follows easily.

Although we have fixed the write positions $i_1,\ldots,i_k$, we will consider
all the possible written values $\vec{a} = \langle a_1,\ldots,a_k\rangle$. Let 
$M(\vec{a})$ be the state of the memory after all the $\writ$ operations are 
carried out, and $M^i(\vec{a})$ be the state $M(\vec{a})$ where all the cells
with stamp $i$ are restored to the value they had before epoch $i$ started.
Let also $Q(\vec{a}) = \langle \prefix(B,1),\ldots,\prefix(B,n)\rangle$ be 
\new{the answers to all the possible queries} 
after the upates of $\vec{a}$ are performed (i.e., run on $M(\vec{a})$), and 
$Q^i(\vec{a})$ be the same vector of answers when running on $M^i(\vec{a})$.
Consider now the following inequalities:
\begin{eqnarray}
&& \textrm{Worst-case complexity of $\prefix$} \nonumber \\
&\ge & \max_{\vec{a} \in \{0,1\}^k, y \in [1\dd n]} \textrm{time of
$\prefix(B,y)$ on $M(\vec{a})$} \label{eq:1} \\
&\ge & \frac{1}{2^k n} \sum_{\vec{a}\in\{0,1\}^k} 
			\sum_{y \in [1\dd n]} \textrm{time of
$\prefix(B,y)$ on $M(\vec{a})$} \label{eq:2} \\
&\ge& \frac{1}{2^k n} \sum_{\vec{a}\in\{0,1\}^k}
                        \sum_{y \in [1\dd n]} 
			\sum_{i=1}^r [\textrm{$\prefix(B,y)$ reads some cell
stamped $i$ in $M(\vec{a})$]} \label{eq:3} \\
&= & \frac{1}{2^k n}  \sum_{i=1}^r \sum_{\vec{a}\in\{0,1\}^k} 
        |\{ y, \textrm{$\prefix(B,y)$ reads some cell stamped $i$ in
$M(\vec{a})$}\}| 
\label{eq:4}\\
&\ge& \frac{1}{2^k n} \sum_{i=1}^r \sum_{\vec{a}\in\{0,1\}^k}
			\mathbf{d}(Q(\vec{a}),Q^i(\vec{a})) \label{eq:5}
\end{eqnarray}
Formula~(\ref{eq:1}) follows by definition (it is an inequality because we
have fixed, for example, the $\writ$ positions), 
and it upper bounds (\ref{eq:2}),
which is the average cost of $\prefix$. This upper bounds (\ref{eq:3}), where
we charge only one unit of work per epoch stamp $\prefix$ reads (the notation
$[p]$ means $1$ if predicate $p$ holds and $0$ if not). We just reorganize
terms in (\ref{eq:4}). The term $\mathbf{d}$ in (\ref{eq:5}) refers to the
Hamming distance between the two binary vectors. The inequality holds because,
if an answer changes between $Q^i(\vec{a})$ and $Q(\vec{a})$, this means that 
$\prefix$ must have read some cell that was written in epoch $i$.
We now split $\vec{a} = \vec{a_1} \cdot \vec{a_2}$, where $\vec{a_2}$ contains
the older $l_i$ updates and $\vec{a_1}$ the newer $k-l_i$ ones.
\begin{eqnarray}
&& \frac{1}{2^k n} \sum_{i=1}^r \sum_{\vec{a}\in\{0,1\}^k}
			\mathbf{d}(Q(\vec{a}),Q^i(\vec{a})) \nonumber \\
&=& \frac{1}{2^k n} \sum_{i=1}^r \sum_{\vec{a_1}\in\{0,1\}^{k-l_i}}
		      \sum_{\vec{a_2}\in\{0,1\}^{l_i}}
			\mathbf{d}(Q(\vec{a_1}\cdot\vec{a_2}),Q^i(\vec{a_1}\cdot\vec{a_2})) \label{eq:6} \\
&\ge& \frac{1}{2^k n} \sum_{i=1}^r \sum_{\vec{a_1}\in\{0,1\}^{k-l_i}}
		\frac{n}{12} \cdot |\{\vec{a_2}\in\{0,1\}^{l_i}, 
			\mathbf{d}(Q(\vec{a_1}\cdot\vec{a_2}),Q^i(\vec{a_1}\cdot\vec{a_2}))
\ge n/12 \}| \label{eq:7} \\
&=& \frac{1}{12 \cdot 2^k} \sum_{i=1}^r \sum_{\vec{a_1}\in\{0,1\}^{k-l_i}}
		 \!\!\!\!\!\!(2^{l_i} - |\{\vec{a_2}\in\{0,1\}^{l_i}, 
			\mathbf{d}(Q(\vec{a_1}\cdot\vec{a_2}),Q^i(\vec{a_1}\cdot\vec{a_2}))
< n/12 \}|) \label{eq:8} \\
&\ge& \frac{1}{12 \cdot 2^k} \sum_{i=1}^r \sum_{\vec{a_1}\in\{0,1\}^{k-l_i}}
		 \!\!\!\!\!\!(2^{l_i} - |A^i|\cdot \textrm{Ham}(A,n/12)) \label{eq:9} 
\end{eqnarray}
In Formula~(\ref{eq:6}) we just decomposed $\vec{a}$, and then applied a simple
convenient lower bounding to obtain (\ref{eq:7}), which is rewritten in
(\ref{eq:8}). In Formula~(\ref{eq:9}), we are defining 
$A=\{ Q(\vec{a_1}\cdot\vec{a_2}), \vec{a_2} \in \{0,1\}^{l_i}\}$,
$A^i=\{ Q^i(\vec{a_1}\cdot\vec{a_2}), \vec{a_2} \in \{0,1\}^{l_i}\}$, and
Ham$(A,n/12)$ as the maximum number of elements of $A$ that can be placed
in a Hamming ball of radius $n/12$. That is, we upper-bound how many vectors 
of $Q(\vec{a_1}\cdot\vec{a_2})$ can there be close enough to each vector 
of $Q^i(\vec{a_1}\cdot\vec{a_2})$. In turn, by the triangle inequality, 
Ham$(A,n/12)$ can be upper-bounded by the elements of $A$ that fall in a 
Hamming ball of twice the radius, $n/6$, centered around \new{some element 
$\vec{c} \in A$ (one that is in the original Hamming ball)}. 
We now bound this value.

The elements of $A$ are correct answer vectors after all the writes take
place. Per our partition $\vec{a_1} \cdot \vec{a_2}$, where we fixed
$\vec{a_1}$, only the last $l_i$ of those updates differ between any two
elements $\vec{c},\vec{v} \in A$. Let $j$ and $j'$ be two writing
positions in $\{ i_1,\ldots,i_{l_i}\}$ that are consecutive in $B$. 
Consider the segments of answers
$c_j c_{j+1}\cdots c_{j'-1}$ of $\vec{c}$ and 
$v_j v_{j+1}\cdots v_{j'-1}$ of $\vec{v}$. This is the key point:
because there are no updates in $B[j+1 \dd j'-1]$, either the parities of
the $\rank$s (i.e., the answers to $\prefix$) between $\vec{c}$ and $\vec{v}$
are all the same along the range, or they are all different. That is,
it must hold either
$v_j v_{j+1}\cdots v_{j'-1} = c_j c_{j+1}\cdots c_{j'-1}$ or
$v_j v_{j+1}\cdots v_{j'-1} = \overline{c_j c_{j+1}\cdots c_{j'-1}}$, where 
$\overline{c}$ denotes the complement of bit $c$. In the first case, the pair
$(j,j')$ contributes zero to the Hamming distance, but due to the minimum
distance between two positions in epoch $i$, the second case contributes at 
least $n/(2l_i)$. Within a ball of radius $n/6$ centered at $\vec{c}$, then, 
there can be at most $l_i/3$ pairs $(j,j')$ that produce changes in the
answers. The number of elements of $A$ within 
the Hamming ball centered at $\vec{c}$ is then bounded by the number of ways
to choose up to $l_i/3$ consecutive pairs $(j,j')$, that is,
$\sum_{t=0}^{\lfloor l_i/3\rfloor} {l_i \choose \new{t}}$. 
This is at most $2^{H(1/3)\,l_i} \le 2^{0.92\,l_i}$, where
$H(p)=p\log_2\frac{1}{p}+(1-p)\log_2\frac{1}{1-p}$ is the binary entropy
function \cite[p.~427]{FG06}.

We now bound $|A^i|$. The number of cells referenced by any $\writ$ operation,
if it performs $t$ units of work in the worst case, can be upper-bounded by 
$2^{wt}$ via the ``decision assignment tree'' model of a deterministic
algorithm: the root indicates the first cell read by the algorithm. It has
$2^w$ children, one per possible content of the cell, by which the execution can
continue. The edge to the child is annotated with the value the algorithm
rewrites on the cell. Then the algorithm reads a second cell, the one
indicated in the child node, which can have $2^w$ possible outcomes, and so on.
After $t$ steps, the number of cells possibly accessed by the algorithm is at 
most $2^{wt}$. The total number of cells accessed along the 
$k$ $\writ$ operations is then $s \le k2^{wt}$.
Therefore,
$$ |A^i| ~=~ |\{ Q^i(\vec{a_1} \cdot \vec{a_2}),
\vec{a_2}\in\{0,1\}^{l_i}\}| ~~\le~~
|\{ M^i(\vec{a_1}\cdot\vec{a_2}), \vec{a_2}\in\{0,1\}^{l_i}\}|,$$
because the number of distinct query result vectors on $M^i(\vec{a})$ is bounded by
the number of distinct memory configurations on which the queries were executed.
We can bound the number of those configurations as
$$|\{ M^i(\vec{a_1}\cdot\vec{a_2}), \vec{a_2}\in\{0,1\}^{l_i}\}| ~~\le~~
    \sum_{j=0}^{t\cdot l_{i-1}} {s \choose j} 2^{wj}
    ~~\le~~ \sum_{j=0}^{t\cdot l_{i-1}} {k2^{wt} \choose j} 2^{wj}.$$
This is because, once $\vec{a_1}$ is fixed, all the changes in epoch $i$ are
restored to their original value in $M^i(\vec{a})$. The changes that remain are
those
applied on the epochs $i-1$ to $1$, on which we perform $l_{i-1}$ $\writ$
operations. Those $l_{i-1}$ operations may affect $j$ cells, with $j$ ranging
from $0$ to $t \cdot l_{i-1}$. For each possible set of $j$ cells, of which 
there are ${s \choose j}$, there are by the discussion above at most 
$2^{wj}$ different executions---and thus different memory configurations.
Now we show that this value is in $2^{o(l_i)}$. Bounding
$$ \sum_{j=0}^{t\cdot l_{i-1}} {k2^{wt} \choose j} 2^{wj} ~~\le~~
(t \cdot l_{i-1}) \cdot (k 2^{wt})^{t\cdot l_{i-1}} \cdot 2^{wt\cdot
l_{i-1}},$$
remembering that
$l_i = \log^{3i} n$, \new{$i \le r = O(\log n/\log\log n)$,} $w = \log n$, 
$k \le n$, and assuming \new{$t = O(\log n/\log\log n)$} (i.e., the optimal 
\new{worst-case} time for the $\writ$ operation),
we take logarithm on the last term to obtain
$$\new{O\left(\log n + (\log^{3i-2} n/\log\log n)(\log n + \log^2 n/\log\log n) +
\log^{3i-1}n/\log\log n\right),}$$
which is $o(\log^{3i} n) = o(l_i)$.
We are now ready to finish our chain of inequalities:
\begin{eqnarray}
&& \frac{1}{12 \cdot 2^k} \sum_{i=1}^r \sum_{\vec{a_1}\in\{0,1\}^{k-l_i}}
		 \!\!\!\!\!\!(2^{l_i} - |A^i|\cdot \textrm{Ham}(A,n/12)) 
\nonumber \\
& \ge & \frac{1}{12 \cdot 2^k} \sum_{i=1}^r \sum_{\vec{a_1}\in\{0,1\}^{k-l_i}}
                 \!\!\!\!\!\!(2^{l_i} - 2^{o(l_i)} \cdot 2^{0.92\,l_i})
\label{eq:10} \\
& \ge & \frac{1}{12 \cdot 2^k} \sum_{i=1}^r \sum_{\vec{a_1}\in\{0,1\}^{k-l_i}}
                 \!\!\!\!\!\!2^{l_i}/2 \label{eq:11} \\
& = & \frac{1}{24 \cdot 2^k} \sum_{i=1}^r \sum_{\vec{a_1}\in\{0,1\}^{k-l_i}}
                 \!\!\!\!\!\!2^{l_i} 
~~=~~ \frac{1}{24 \cdot 2^k} ~ r \cdot 2^k ~~=~~ \frac{r}{24} ~~=~~
\frac{\log(n/q)}{72 \log\log n}.  \label{eq:12} 
\end{eqnarray}
The bounds we have derived justify Formula~(\ref{eq:10}). Then the step to
(\ref{eq:11}) holds for sufficiently large $n$. The final result in
(\ref{eq:12}) follows easily.

\medskip
\begin{theorem} \label{thm:lb}
In the cell probe model, with computer words of size $w=\log n$, \new{some of 
the operations} on a dynamic bitvector of length $n$, where the fraction of 
updates is $1/q$, \new{must require} $\Omega(\log(n/q)/\log\log n)$ \new{time in
the worst case}.
\end{theorem}
\medskip

Note that we have proved a lower bound just on a sequence of operations $\writ$
and $\rank$. The lower bound trivially holds for supersets of those operations,
but we can also obtain uncomparable results by reductions. An easy one
is to reduce $\writ$ to $\insertar$ and $\delete$, so we can replace it by the
two operations in the lower bound. We can also replace $\rank$ by $\select$ in
the lower bound \cite[Lem.~7.3]{RRR07}, 
by representing $B[1\dd n]$ with a bitvector $B'$ of length
up to $2n$ where every $B[i]=0$ is encoded as a $1$ and every $B[i]=1$ is
encoded as $01$. It then holds that $\rank_1(B,i)=\select_1(B',i)-i$ (and
$\select_1(B,i)=\select_0(B',i)-i+1$). We can also reduce to $\select_0$ by
complementing the bits in the encoding. The lower bound, then, holds for 
\new{the operation} $\writ$ (or, alternatively, $\insertar/\delete$) and
{\em any} of the queries $\rank$, $\select_1$, or $\select_0$.

\section{Applications} \label{sec:app}

Navarro \cite[Ch.~12]{Nav16} describes several dynamic compact data structures
that can be built from dynamic bitvectors. We revisit them in this section,
showing how our new result yields adaptive dynamic variants (see the chapter
for more details). An original aspect of this section is the way it combines
static and dynamic versions in a single data structure; this uncovers 
interesting tradeoffs in some cases.

\subsection{Arrays with cells of fixed length}

Instead of a dynamic array of bits, we might have a dynamic array of $\ell$-bit 
elements $A[1\dd n]$, where we wish to perform operations $\insertar(A,i,v)$,
which inserts value $v$ at $A[i]$, $\delete(A,i)$, which removes $A[i]$,
$\leer(A,i)$, which returns $A[i]$, and $\writ(A,i,v)$, which sets $A[i]
\gets v$. We assume $\ell \le w$.

Dynamic fixed-length arrays can be implemented analogously to dynamic 
bitvectors, using $b=\Theta(w\log n/\log\log n)$ bits per leaf and holding up to
$\lfloor b/\ell\rfloor$ cells in each leaf. Allocating an integral number of 
words wastes at most $w$ bits per leaf, which amounts to 
$O(n\ell\log\log n/\log n)$ bits in total. 
Note that operations $\leer$ and $\writ$ on a leaf take constant time, while 
$\insertar$ and $\delete$ take time $O(b/w)=O(\log n/\log\log n)$, as 
we move the data by whole computer words. Because of the larger leaf sizes we
can afford, we can use $w$-bit system pointers at the nodes.

We can apply the same techniques developed for adaptive dynamic bitvectors, so
that we flatten nodes that are updated infrequently. While a dynamic leaf can 
also be read in constant time, a flattened leaf has smaller depth and thus 
fewer tree nodes are traversed to reach it. Interestingly, we should not
consider $\writ$ to be an update operation, as it can be performed on static
leaves in constant time as well.

\medskip
\begin{corollary}
An adaptive dynamic array of $\ell$-bits cells, starting empty, can be 
maintained in $n\ell(1+o(1))$ bits of space,
where $n$ is the current number of cells it represents, so that if the
fraction of insertions/deletions over total operations so far is $1/q$, 
then the array operations take $O(\log(n/q)/\log\log n)$ amortized time.
\end{corollary}
\medskip

\no{
\textcolor{blue}{Our public code includes an implementation of this data structure, together 
with that of bitvectors.}
}

\subsection{Arrays with cells of variable length}

In a more complex scenario, we want to store an array $A[1\dd n]$
where each cell has a different length, which is usually inferred from the
cell content (i.e., each cell contains a self-delimiting variable-length code
like Huffman, $\gamma$-codes, $\delta$-codes, etc.~\cite{Sol07}).
Let $N$ be the sum of all the lengths of the cells in $A$. We can represent
$A$ as a bitvector $B[1\dd N]$ that concatenates all the cells of $A$. 

Our dynamic representation of $B$ stores an integral number of cells of $A$
in each leaf; we assume again that cell lengths do not exceed $w$ 
bits. We return to using leaves of $b=\Theta(\log^2 n/\log\log n)$ bits,%
\footnote{In the awkward case where a $w$-bit cell does not fit within a leaf,
we can allocate a $w$-bit leaf for that cell.}
so the wasted space is $O(N/\log n)$ bits. The total space incurred by our data 
structure is then within $N+o(N)$ bits.

Processing a leaf, for queries or for updates, requires scanning it
to decode its cells. If cells are short, this may require decoding up to  
$O(\log^2 n/\log\log n)$ individual cells. We can use precomputed tables of size
$o(n)$ to decode $\Theta(\log n)$ bits (which may encode several cells) in 
constant time, analogous to those used in Section~\ref{sec:adb};
this mechanism has been detailed, for example,
for $\delta$-codes \cite[Sec.~5.1.1]{MN08}. With this method, cells of 
length $\Omega(\log n)$ are still processed one by one, so we assume each 
cell can be decoded in constant time. When we execute $\writ$, we must also
encode one cell, for which we may allow $O(\log n/\log\log n)$ time; the rest of the cost
is for copying memory in the leaf, which can be done by chunks of $w$ bits. 
Overall, leaves can be processed in $O(\log n/\log\log n)$ time under these assumptions.
We note that operation $\writ(A,i,v)$ might change the length of cell $A[i]$, 
and therefore it must be treated as an update just like $\insertar$ and 
$\delete$, as it requires shifting bits in, and even resize, the (dynamic) leaf.

For static leaves $v$, let $B_v$ be the local piece of the bitvector $B$
stored at $v$. We can use a sparse static bitvector $S_v$ 
\cite{Eli74,Fan71,OS07} to mark the starting positions of the entries in
$B_v$. To perform $\leer(B_v,i)$, we extract the bits from the positions 
$\select_1(S_v,i)$ to $\select_1(S_v,i+1)-1$ of $B_v$; this takes constant
time.
Such representation uses $n_i\log_2(N_i/n_i)+O(n_i)$ bits for the $i$th static 
leaf, storing $n_i$ cells of total length $N_i$. Added over all the static
leaves we get $\sum_i n_i\log_2(N_i/n_i)+O(n_i) \le n\log_2(N/n) + O(n)$ 
extra bits of space, using Jensen's inequality. This is $o(N)+O(n)$ bits.

\medskip
\begin{corollary} \label{cor:varlength}
An adaptive dynamic array of variable-length cells, starting empty, 
can be maintained in $N+o(N)+O(n)$ bits of space, where $n$ is the current 
number of cells and $N$ the sum of the lengths of the cells,
so that if the fraction of updates over total operations so far is $1/q$, 
then the array operations take $O(\log(n/q)/\log\log n)$ amortized time. 
This assumes cells can be decoded in constant time and encoded in 
$O(\log n/\log\log n)$ time.
\end{corollary}

\paragraph{Compressed bitvectors}

Dynamic compressed bitvectors \cite{BB04,MN08,NS14} aim to represent bitvector 
$B[1\dd n]$ using $nH$ bits of space; the bitvector entropy $0 \le H \le 1$ was
defined in Section~\ref{sec:related}. A (static) representation of $B$ within 
$nH + o(n)$ bits \cite{RRR07}, which is built in linear time, 
divides it into chunks of $\kappa=(\log_2 n)/2$ bits, and 
stores for each chunk its ``class'' $c$ (number of 1s) in $\lceil \log_2 
(\kappa+1)\rceil$ bits, and its ``offset'' $o$ (index within the class) using 
$\lceil \log_2 {\kappa \choose c} \rceil$ bits. While the $c$ components add
up to $o(n)$ bits, it is shown that the lengths of the $o$ components adds up
to $nH$, which still holds if we distribute them across leaves.

While the classes $c$ can be stored in fixed-length cells, the offsets $o$
have variable length, which can be known from their class $c$. The 
static representation can be used for the static leaves, as it offers
$\access$, $\rank$, and $\select$ in constant time.
On the dynamic leaves, we store an integral number of chunks
and implement the operations via scanning. We use again precomputed tables, as 
already described in the literature \cite[Sec.~5.2.1]{MN08}, to obtain 
$O(\log n/\log\log n)$ scanning time. 
The operations $\insertar$ and $\delete$ at the bit level require re-encoding 
all the chunks of the leaf that lie to the right of the affected one, which is 
also done by chunks of $\Theta(\log n)$ bits \cite[Sec.~5.2.2]{MN08}. A variant of 
the (class,offset) encoding ensures that one insertion or deletion makes the
size of the leaf grow or shrink by $O(\log n)$ bits \cite[Sec.~8.2]{NS14}.

\medskip
\begin{corollary} \label{cor:rrr}
An adaptive dynamic bitvector starting empty can be maintained in 
$nH+o(n)$ bits of space, where $n$ is the current number of bits and $H$ its 
entropy, so that if the fraction of updates over total operations so far is 
$1/q$, then the bitvector operations take $O(\log(n/q)/\log\log n)$ 
amortized time.
\end{corollary}

\paragraph{Very sparse bitvectors}

An alternative formula of entropy compression is $nH = m\log_2(n/m)+O(m)$,
where $m$ is the number of 1s in $B$. This shows that the $o(n)$ bits of
redundancy in the previous scheme is too large when $m \ll n$. A representation
that avoids that redundancy stores, in a variable-length array of $m$ elements, 
the lengths of the gaps between consecutive 1s in $B$. If we encode those values
using $\delta$-codes, then the sum $N$ of the cell lengths is upper 
bounded as $N \le m\log_2(n/m) + O(m\log\log(n/m)) = nH + o(nH) + O(m)$ bits. As
before, this also holds if we distribute the sequence across several leaves.

We can use, in principle, Corollary~\ref{cor:varlength} to represent this 
sequence. Operations $\insertar$ and $\delete$ on $B$ translate into $\writ$,
$\insertar$ and $\delete$ on the sequence. As for queries, the sequence 
representation enables $\leer$ operations on the cells, but we 
need another functionality on $B$: $\access$, $\rank$, and $\select$.
On dynamic leaves, we can implement them by
scanning the sequence of gap lengths. This can be done by chunks of
$\Theta(\log n)$ bits so that leaves are processed in $O(\log n/\log\log n)$ 
time \cite[Sec.~5.1.1]{MN08}. 

On the static leaves, we do not use the $O(n\log(N/n))$-bit sized 
bitvectors $S_v$; we directly encode $B_v$ using the representation we 
chose for $S_v$ \cite{OS07}. Such representation supports 
$\select_1(B_v,j)$ in $O(1)$ time and the other queries in time 
$O(\log m)$. 

Let the $i$th $v$ leaf store a bitvector of length $n_i$ with $m_i$ 1s. If the
$i$th leaf is dynamic, then it is represented with $\delta$-codes and its space
is $m_i\log_2(n_i/m_i) + O(m_i\log\log(n_i/m_i))$, as already said. If it is
dynamic, the space of its representation is $m_i\log_2(n_i/m_i) + O(m_i)$
\cite{OS07}. The sum can then be bounded as 
$\sum_i m_i\log_2(n_i/m_i) + O(m_i\log\log(n_i/m_i)) \le
m\log_2(n/m)+O(m\log\log(n/m)) = nH + o(nH) + O(m)$ bits, using Jensen's
inequality. Our representation then uses the same asymptotic space.

We note that the $O(\log m)$ time for $\rank$ and $\select_0$ is particularly 
high in this scenario; they run faster on dynamic leaves (i.e., $O(\log
m/\log\log m)$ time) than on static ones! A solution is to encode, in static
leaves, a static version of our WBB-tree, which instead of explicitly storing
the leaves points to their position in the static bitvector. Such static
WBB-trees add only $o(nH)+O(m)$ further bits and reduce the times as
described. Lower times for $\rank$ can be obtained with other static
representations \cite{GHSV07}.

\medskip
\begin{corollary} \label{cor:verysparse}
An adaptive dynamic bitvector starting empty can be maintained in 
$nH+o(nH)+O(m)$ bits of space, where $n$ is the current number of
bits, $H$ is its entropy, and $m$ is its number of 1s, 
so that if the fraction of updates over total operations so far is $1/q$, 
then the bitvector operations $\insertar$, $\delete$, and $\select_1$ have 
$O(\log (m/q)/\log\log m)$, and the rest $O(\log m/\log\log m)$, amortized time.
\end{corollary}
\medskip

\paragraph{Searchable partial sums}

A sparse bitvector like that of Corollary~\ref{cor:verysparse}
can be used to implement searchable 
partial sums with indels \cite{HSS03}. Here we aim to represent an
array $A[1\dd n]$ of positive numbers so that we can, in addition, support the
operations $\suma(A,i)=\sum_{j=1}^i A[j]$ and $\search(A,v) = \max 
\{ i,~\suma(A,i) \le v\}$. If we let $N=\suma(A,n)$ and represent $A$ as a
bitvector $B[1\dd N]$, where we set a $1$ at every position $\suma(A,i)$, then
it follows that $\suma(A,i)=\select_1(B,i)$, $\search(A,v)=\rank_1(B,v)$, and
$\leer(A,i)=\select_1(B,i)-\select_1(B,i-1)$. 

By using Corollary~\ref{cor:verysparse}, we would actually be representing the
entries of $A$ in variable-length cells, using $\delta$-codes in the dynamic
leaves. Operations $\writ$, $\insertar$, and $\delete$ on $A$ boil down to
similar operations on the $\delta$-codes. A problem, however, is that the
internal array $v.\size$ (and also $v.\zeros$, but we do not need that one
here) may change by arbitrarily large amounts, which is not supported in the
data structure we use to maintain the array \cite{RRR01}. This is a
fundamental problem, as otherwise we would break lower bounds \cite{PD06}.

We can obtain logarithmic times by using a constant tree arity $a$ in our data
structure. A consequence is that, if we suspend flattening on a constant
number of levels, we can only ensure we flatten nodes of size at most
$\epsilon \cdot n$ for some constant $\epsilon>0$; therefore the space grows
by a constant fraction. We then have the following result.

\medskip
\begin{corollary}
An adaptive searchable partial sum with indels, starting empty, can be 
maintained in $(1+\epsilon)n\log_2(N/n)+O(n)$ bits of space, for any constant
$\epsilon>0$, where $n$ is the current number of elements and $N$ their sum,
so that if the fraction of updates over total operations so far is $1/q$, 
then the operations $\insertar$, $\delete$, $\leer$, $\writ$, and
$\suma$ have $O(\log (n/q))$, and $\search$ has $O(\log n)$, amortized time.
\end{corollary}

\subsection{Wavelet trees and matrices}

Wavelet trees \cite{GGV03} represent a sequence $S[1\dd n]$ over alphabet
$[1\dd\sigma]$ so that various operations can be carried out on it, in
particular $\access(S,i)$, which yields $S[i]$, $\rank_c(S,i)$, which gives
the number of times $c$ occurs in $S[1\dd i]$, and $\select_c(S,j)$, which is
the position of the $j$th occurrence of $c$ in $S$. The wavelet tree is a
balanced tree of $O(\sigma)$ nodes storing bitvectors at every node, adding up 
to $n$ bits per level. Its operations are carried out in $O(\log\sigma)$ time 
by reducing them to $O(\log\sigma)$ operations on the bitvectors, one per
level. Dynamic representations \cite{NS14} support the operations $\insertar$ 
and $\delete$ on $S$ and obtain time $O(\log\sigma \log n / \log\log n)$ per 
operation, and even $O(\lceil \log\sigma/\log\log n\rceil \log n/\log\log n)$
by reducing the wavelet tree to small subalphabets instead of to bits (this can
probably be made adaptive too, but here we stick to the case of bits). 
A wavelet matrix \cite{CNO15} is formed by
$\log_2\sigma$ bitvectors of length $n$ (one per wavelet tree level) and 
simulates the same operations of the wavelet
tree without spending $O(\sigma \log n)$ bits to store the nodes. 

We note that each query/update on the wavelet tree or matrix translates into one
query/update on each level of the bitvectors. Thus the number $q$ of queries 
per update stays the same over the bitvectors of every level. Given $m$ wavelet
tree operations, $u$ of the operations being updates, with $q=m/u$, let us
consider how the (same number of) operations distribute along the bitvectors 
of a given level. Let the $i$th bitvector, of length $n_i$, receive $m_i$ 
operations of which $u_i$ are updates. The total amortized time using our 
adaptive dynamic bitvectors is then $\sum_i m_i\log(n_i/q_i)$ divided by 
$O(\log\log n)$, where $q_i=m_i/u_i$, $\sum_i m_i = m$, 
$\sum_i n_i = n$, and $\sum u_i = u$. By 
Jensen's inequality, the sum is at most $m \log \sum_i m_i n_i / (q_i m)
= m \log \sum_i (n_i u_i)/m \le 
m \log (\sum_i n_i) (\sum_i u_i)/m = 
m\log (nu/m) = m\log(n/q)$.
We then obtain the following.

\medskip
\begin{corollary} \label{cor:wt}
An adaptive dynamic wavelet tree or matrix over alphabet $[1\dd\sigma]$, 
starting empty, can be maintained in $n\log_2 \sigma + o(n\log\sigma)$ bits of 
space (plus $O(\sigma \log n)$ bits in the case of a wavelet tree), where $n$ 
is the current number of sequence elements, so that if the 
fraction of updates over total operations so far is $1/q$, then the operations 
have $O(\log\sigma \log (n/q)/\log\log n)$ amortized time.
\end{corollary}
\medskip

We note that there exist sequence representations that carry out all the
operations in time $O(\log n/\log\log n)$ \cite{MN15}; this is incomparable
with ours. By combining wavelet trees with compressed bitvectors, giving them
Huffman shape, and using many other techniques, one can obtain a wide range of
space-time tradeoffs for sequences \cite{Nav14}; they all have their
corresponding counterparts if combined with our adaptive dynamic bitvectors.
We leave exploring those as an exercise to the reader. Instead, we will focus
on other data structures that can be implemented on top of wavelet trees or
matrices.

\paragraph{Discrete grids}

A discrete grid of $r$ rows and $c$ columns contains $n$ points at positions
$(i,j) \in [1\dd r] \times [1\dd c]$, and supports queries like counting how 
many points are there in a rectangle $[r_1 \dd r_2] \times [c_1 \dd c_2]$, or 
reporting those points, among others. If the grid is dynamic, we can also 
insert and delete points at any grid position. A way to maintain a grid is by
combining a bitvector $B[1\dd c+n] = 1 0^{n_1} 1 0^{n_2} \cdots 1 0^{n_c}$,
which signals that there are $n_j$ points in column $j$, with a wavelet matrix
$S[1\dd n]$, which gives the row coordinates of the points, read in increasing
column order. This uses $(c+n\log_2 r)(1+o(1))$ bits.

Queries on the grid are translated to queries on the sequence using $\select$ 
on $B$, for example the column range $[c_1\dd c_2]$ becomes the
string range $S[\select_1(B,c_1)-c_1+1 \dd \select_1(B,c_2+1)-(c_2+1)]$. On
the other hand, a string position $S[i]$ corresponds to the point
$(\select_0(B,i)-i,S[i])$, if we want to report it. Those conversions take
constant time in the static case. Further, the operation that counts
the number of points on a rectangle is implemented in time $O(\log r)$ using
the wavelet matrix, whereas each point is reported in time $O(\log r)$ as
well. In the dynamic structure, inserting/deleting a point corresponds to
inserting/deleting a $0$ in $B$ and a symbol in the wavelet matrix. By using
our adaptive bitvector representation, we obtain the following result.
 
\medskip
\begin{corollary}
An adaptive dynamic grid of $r$ rows and $c$ columns, starting empty, can be 
maintained in $(c+n\log_2 r)(1+o(1))$ bits of space,
where $n$ is the current number of points, so that if the 
fraction of updates over total operations so far is $1/q$, then points can be
inserted and deleted, the points within a rectangle can be counted, and each
such point can be retrieved, in $O((\log(c/q)+\log (n/q)\log r)/\log\log n)$ 
amortized time.
\end{corollary}
\medskip

We note that each of the reported points counts as a query to the bitvectors, 
so the values of $q$ are high if we use reporting queries.
Other operations that take $O(\log r)$ time are similarly translated to the
adaptive dynamic case.

\paragraph{Graphs}

The same data structure can be used to represent a directed graph of $n$ nodes 
and $e$ edges: the $n \times n$ grid has a point in $(i,j)$ iff there is an
edge from node $i$ to node $j$. Insertions and deletions of edges correspond
to insertions and deletions of points in the grid, and adjacency queries,
queries for all the neighbors of a node, or all its reverse neighbors, are
translated to rectangle queries on the grid.

\medskip
\begin{corollary}
An adaptive dynamic graph on $n$ nodes, starting empty of edges, can be 
maintained in $(n+e\log_2 n)(1+o(1))$ bits of space,
where $e$ is the current number of edges, so that if the 
fraction of updates over total operations so far is $1/q$, then edges can be
inserted and deleted, an adjacency can be queried, and the neighbors and
reverse neighbors of a node can be counted and each can be enumerated,
in $O(\log (e/q)\log n/\log\log e)$ amortized time.
\end{corollary}

\paragraph{Texts}

Our final application of wavelet trees is to maintain a dynamic collection of
text documents, so that we can insert and delete whole documents, and search for
patterns on those \cite{CHL04,MN08,NS14}. The main data structure we maintain
is the Burrows-Wheeler Transform (BWT) \cite{BW94} of the concatenation of the
texts. The BWT is a permutation of the symbols of the collection in an order
that is suitable for compression and for indexed searching \cite{FM05}. The
search for a short pattern $P[1\dd m]$ is done in $O(m\log\sigma)$ time,
corresponding to $O(m)$ operations on the wavelet tree of the BWT. After this
time, one can tell how many times $P$ occurs in the collection, and then
can output the text position of each such occurrence in time $O(l \log \sigma)$,
 where $l$ is a sampling step that induces $O(l \log n)$ extra bits of space,
$n$ being the total length of the document collection.
If the wavelet tree compresses the bitvectors with the technique we used in
Corollary~\ref{cor:rrr} \cite{RRR07}, then its total space is 
$nH_k+o(n\log\sigma)$ bits for any $k \le \alpha \log_\sigma n$ and constant
$\alpha < 1$, where $H_k$ is the $k$th order
empirical entropy of the collection \cite{MN08}.

The insertion of a new text $T[1\dd n']$ into the collection, or the deletion
of $T[1\dd n']$ from the collection, requires $O(n')$ queries and updates on 
the wavelet tree of the BWT. In addition, dynamic arrays with fixed-cell width 
of $O(n/l)$ entries must be maintained. By using Corollaries~\ref{cor:wt} and
\ref{cor:varlength}, we obtain the following result.

\medskip
\begin{corollary}
An adaptive dynamic text collection on alphabet $[1\dd\sigma]$, starting empty,
can be maintained in $nH_k + o(n\log\sigma) + O((n\log n)/l)$ bits of space, 
where $n$ is the current size of the 
collection and $l$ is a sampling step, so that if the fraction of updates over 
total operations so far is $1/q$, then the occurrences of a pattern $P[1\dd m]$
can be counted in time $O(m\log\sigma\log(n/q)/\log\log n)$, each of its 
occurrences can be located in time $O(l\log\sigma\log(n/q)/\log\log n)$, and 
any text of length $n'$ can be inserted in or deleted from the collection in 
$O(n'\log\sigma\log(n/q)/\log\log n)$ time; all times are amortized.
\end{corollary}
\medskip

Again, we note that a pattern search counts as $O(m)$ queries, a locate as
$O(l)$ queries, and an insertion/deletion of a document as $O(n')$ updates. 
This is a case
where updates are bursty, as we perform the $O(n')$ update operations
together.

\subsection{Trees}

A simple representation of ordinal trees of $n$ nodes, which fits particularly 
well with our result, is LOUDS \cite{Jac89}. This representation consists of a
bitvector $B[1\dd 2n]$, which is built by traversing the tree levelwise
starting from the root, left to right on each level, and at each node appending
its ``signature'' $1^c 0$ to $B$, where $c$ is the number of children of
the node. Nodes are identified with the position where their signature
starts. Navigation of the static representation can be performed in constant
time: the number of children of node $v$ is $\degree(v) = \next_0(B,v)-v$, its 
$i$th child is $\child(v,i)=\select_0(B,\rank_1(B,v-1+i))+1$, its parent is 
$\parent(v)=1+\prev_0(B,j)$, where $j=\select_1(B,\rank_0(B,v-1))$, and the
position of $v$ among the children of its parent is $\childrank(v) =
j-\parent(v)+1$. Here
$\prev_0$ and $\next_0$ look for the closest preceding or following $0$; they
can be implemented with $\rank$ and $\select$ or more directly.

A dynamic LOUDS representation uses a dynamic bitvector representation for
$B$. It can insert a new leaf child of $v$ at position $i$ with 
$\insertar(B,v,1)$ and then $\insertar(B,\child(v,i),0)$. Analogously, it can
delete a leaf $v$ by computing $j$ as above and then doing $\delete(B,v)$ and
then $\delete(B,j)$. By using our adaptive dynamic bitvector representation, we
obtain the following result.

\medskip
\begin{corollary}
An adaptive dynamic ordinal tree, starting empty, can be maintained in 
$2n+o(n)$ bits of space, where $n$ is the 
current number of tree nodes, so that if the fraction of updates over total 
operations so far is $1/q$, then all the LOUDS navigation operations, as well
as insertions and deletions of leaves, can be performed in
$O(\log(n/q)/\log\log n)$ amortized time.
\end{corollary}

\paragraph{Cardinal trees}

Unlike ordinal trees, cardinal trees have their children labeled in
$[1\dd\sigma]$, with at most one child per label. An example are binary
trees (with $\sigma=2$). A LOUDS-like representation of cardinal trees
performs a levelwise traversal and writes a $\sigma$-bit signature, with 1s
at the positions for which children exist. Such LOUDS-based cardinal tree
representation uses $\sigma n$ bits, and it can be navigated with operations
$\rank$ and $\select$, much as in the way we have described for standard LOUDS. 

Inserting a leaf into such an ordinal tree requires setting to $1$ the
corresponding position of the parent of the node, and then inserting a block
of $\sigma$ 0s at the corresponding child position. Analogously, deleting a
leaf requires removing the block of $\sigma$ 0s, and then setting to $0$ its
position in its parent. 
For these operations to work smoothly in our scheme, it is best to
consider $B[1\dd \sigma n]$ as an array $A[1\dd n]$ of cells of fixed length
$\sigma$. This array must be enriched so that it carries out operations
$\rank$ and $\select$ over the underlying bitvector, so as to support
navigation. It is not hard to combine
both functionalities to obtain the following result.

\medskip
\begin{corollary} \label{cor:cardinal}
An adaptive dynamic cardinal tree of constant arity $\sigma$, starting empty, 
can be maintained in $\sigma n+o(\sigma n)$ bits of space,
where $n$ is the current number of tree nodes, so that if the 
fraction of updates over total operations so far is $1/q$, then all the
LOUDS-based navigation operations, as well as insertions and deletions of 
leaves, can be performed in $O(\log(n/q)/\log\log n)$ amortized time.
\end{corollary}
\medskip

A particular kind of cardinal trees are the $k^2$-trees \cite{BLN13}, which
are cardinal trees of arity $\sigma=k^2$ that represent the recursive 
partitioning of an $\ell \times \ell$ grid into $k^2$ subgrids. The maximum 
depth of a leaf in the $k^2$-tree is $\lceil \log_k \ell\rceil$, so one does 
not indicate that last-level nodes are leaves (thereby saving the storage of 
those $k^2$ 0s). If storing $n$ points, the most basic form of $k^2$-trees 
requires at most $k^2 n \log_k \ell$ bits. The $k^2$-tree
can determine whether a cell contains a point in time $O(\log_k \ell)$, and
the dynamic variant needs $O(\log_k\ell)$ updates to the bitvector to add or
remove points from the grid, because it may have to insert or delete a full
path of $\log_k \ell$ nodes.

These trees use less space when the grid is clustered, and have
been successfully used to represent web graphs, social networks, grids, and
many other structures. Their performance when implemented using
Corollary~\ref{cor:cardinal} is as follows.

\medskip
\begin{corollary}
An adaptive dynamic $k^2$-tree on an $\ell \times \ell$ grid, starting empty, 
can be maintained in $k^2 n\log_k \ell (1+o(1))$ bits of space,
where $n$ is the current number of points in the grid, so that 
if the fraction of updates over total operations so far is $1/q$, then
accessing cells, as well as inserting and deleting points, can be performed
in $O(\log_k n \log(n/q)/\log\log n)$ amortized time.
\end{corollary}
\medskip

\paragraph{Parentheses}

Another popular representation of ordinal trees uses a sequence of $2n$ 
balanced parentheses \cite{MR01}. The sequence is built by traversing the
tree in depth-first order, appending a `(' when first arriving at a node 
and a `)' when finally leaving it. The sequence is then regarded as a
bitvector $B[1\dd 2n]$. A so-called
fully-functional (static) representation \cite{NS14} is built in linear time and
supports a large number of operations in constant time by navigating the 
parentheses sequence.

Such a representation is advantageous for dynamism: by inserting a couple of
matching parentheses at the correct positions, we can represent the insertion
not only of leaves, but of nodes in the middle of an edge, and in general
of a node $u$ that becomes a child of a node $v$ and replaces $v$'s $i$th to 
$j$th children, which now become children of $u$ \cite{NS14}. Similarly, by 
deleting a pair of matching parentheses, we can remove leaves and internal 
nodes $u$, leaving their current children as children of $u$'s parent.

The representation also offers a much richer set of navigation operations
compared to LOUDS: apart from the basic navigation queries supported by LOUDS, 
we can determine the depth, subtree size, height, number of leaves, leaf range,
pre and postorder rank, iterated ancestors, and deepest descendant leaf of 
nodes, as well as ancestorship and lowest common ancestors of node pairs,
among others \cite[Tab.~I]{NS14}. All those operations are
implemented on top of the concept of {\em excess}: the excess of a position
in the parentheses sequence is the number of opening minus closing parentheses
up to that point. As we identify nodes with the position of their opening
parenthesis, the excess is naturally the depth of the node. A few primitives
are built on top of the excess: forward/backward search, which given a position
and a desired excess find the closest following/preceding position in the
sequence having that excess, the position of the minumum/maximum excess in a range, 
and the number of times the minimum excess occurs in a range.

A simple solution supporting those primitives is the range min-Max tree
(rmM-tree) \cite{NS14}, which cuts the bitvector into blocks and builds a
perfect binary tree on top of them, so that each rmM-tree node represents a
range of the bitvector. Each such node stores a few fields: total, minimum and
maximum relative excess, and number of excess minima in the bitvector range. 
Any query is then solved in $O(\log n)$ time by partially scanning an initial 
and a final bitvector block, plus traversing an upward and downward path of the
rmM-tree. They \cite{NS14} then show how to speed up all the times to
constant, by using multiary rmM-trees that handle polylogarithmic-sized chunks
of the bitvector and classic data structures that solve inter-chunk queries.

Dynamic versions of this structure \cite{NS14,JR12} identify the rmM-tree with
a balanced binary tree, whose leaves store the bitvector blocks. Parentheses
insertions and deletions easily adjust the fields stored at internal nodes as
they return from the recursion. Overall, they obtain $O(\log n)$ time for all
the queries and node insertions/deletions.

It is also possible to obtain $O(\log n / \log\log n)$ time for most queries
and node insertions/deletions, by using a multiary tree whose internal nodes
store sequences of values like 
those we store for $\rank$ and $\select$, now specialized on the excess 
\cite{CHL07,NS14}. Most of the operations require constant time per
internal node traversed, yet some require time $O(\log\log n)$ and thus their
total time is $O(\log n)$; this is the case of the iterated ancestor.
Further,
in order to support updates in time $O(\log n / \log\log n)$, the operations 
$\degree$, $\child$, and $\childrank$ cannot be supported (note that $\child$ 
is not the only operation that lets us traverse the tree downwards; we can use 
instead the more basic operations that get the first/last child and the 
next/previous sibling of a node).

To use our adaptive dynamic bitvectors to store $B$, we aim to implement the
operations in constant time on the static leaves. If we exclude the three
operations mentioned above---$\degree$, $\child$, and $\childrank$---then
it is possible to build the constant-time static data structure \cite{NS14}
on static leaves (we return soon to these three operations).

\medskip
\begin{corollary}
An adaptive dynamic fully-functional ordinal tree, starting
empty, can be maintained in $2n+o(n)$ bits of space,
where $n$ is the current number of tree nodes,
so that if the fraction of updates over total operations so far is $1/q$, 
then all the operations 
\cite[Tab.~I]{NS14} that take $O(\log n/\log\log n)$ time in their ``variant
1'' can be solved in $O(\log (n/q)/\log\log n)$ amortized time, and those
having $O(\log n)$ time can be solved in $O(\log(n/q))$ amortized time.
\end{corollary}
\medskip

Supporting $\degree$, $\child$, and $\childrank$ requires more complex
updates, which take $O(\log\log n)$ time per internal node, and $O(\log n)$
time overall. The same happens with the complexity of the operations
themselves. Further, the constant-time solution for those three operations
\cite{NS14} assumes that the bitvector represents a valid tree. This is not 
the case of our static leaves, which represent arbitrary ranges of the 
parentheses sequence. To handle those operations, we also include on
the static leaves a multiary static rmM-tree that supports all the
operations in time $O(\log\log n)$ \cite{NS14}.

\medskip
\begin{corollary}
An adaptive dynamic fully-functional ordinal tree, starting
empty, can be maintained in $2n+o(n)$ bits of space,
where $n$ is the current number of tree nodes,
so that if the fraction of updates over total operations so far is $1/q$, 
then all the operations 
\cite[Tab.~I]{NS14} that have $O(\log n/\log\log n)$ time in their ``variant
2'' can be solved in $O(\log (n/q)/\log\log n)$ amortized time, and those
having $O(\log n)$ time can be solved in $O(\log(n/q))$ amortized time,
with the exception of $\degree$, $\child$, and $\childrank$, which take
$O(\log(n/q)+\log\log n)$ amortized time.
\end{corollary}
\medskip

A way to obtain $O(\log(n/q)/\log\log n)$ time for those three operations is to 
use parentheses to encode instead the DFUDS representation of the
ordinal tree \cite{BDMRRR05}. In this representation, we traverse the tree in
depth-first order and append to $B$ the LOUDS signature of each
visited node (precisely, $c$ `('s and $1$ `)' if the
node has $c$ children). The sequence turns out to be balanced if we prepend a
`(' to it, and the same primitives we have
discussed can be used to implement most operations, excluding 
the node depth and height, the deepest descendant leaf, postorder
numbering, and iterated ancestors of a node. Another limitation is that
we can insert and delete nodes of constant arity, as we must insert/delete
$c+1$ contiguous parentheses to insert/delete a node with $c$ children. 

In exchange, operations $\degree$, $\child$, and $\childrank$ are supported 
using the most basic primitives, which run in constant time on internal nodes
and static leaves, and hence take $O(\log(n/q)/\log\log n)$ amortized
time on adaptive dynamic bitvectors.

\paragraph{Lower bounds}

Chan et al.~\cite[Thm.~5.2]{CHL07} showed that $\rank$, $\select$, $\insertar$,
and $\delete$ on bitvectors $B[1\dd n]$ can be reduced to the most basic 
problems of maintaining a sequence of balanced parentheses, namely inserting and
deleting pairs of matching parentheses and solving two queries:
\begin{description}
\item[$\match(B,i)$,] the position of the parenthesis matching
that in $B[i]$.
\item[$\enclose(B,i)$,] the position of the opening parenthesis
that most tightly encloses $B[i]$.
\end{description}

The reduction creates a parenthesis sequence $P$ from $B$ as follows: it 
scans $B$ left to right, and for each $B[i]=1$, it appends `((' to $P$; 
if $B[i]=0$ it appends ``()''. After scanning $B$, it appends $r=\rank_1(B,n)$ 
copies of `))' to $P$ so as to make it balanced. The sequence length is then
$m=2n+2r \le 4n$. It is then easy to see that:
\begin{itemize}
\item $\rank_1(B,i) = (m-\match(P,\enclose(P,2i+1))+1)/2$;
\item $\select_1(B,j) = \match(P,m-2j+1)/2$.
\end{itemize}

An update $\insertar(B,i)$ or $\delete(B,i)$ reduces to inserting or deleting 
two parentheses at $P[2i-1]$ and at the end of $P$.
By the lower bound of Fredman and Saks \cite{FS89}, this shows that we need 
$\Omega(\log n/\log\log n)$ time to support insertions and deletions of 
parentheses plus $\match$ and $\enclose$. Per our comments after 
Theorem~\ref{thm:lb}, just supporting updates and $\select$ requires time
$\Omega(\log n/\log\log n)$, and therefore any sequence of operations
consisting of parenthesis insertions and deletions, plus $\match$ queries,
requires time $\Omega(\log n /\log\log n)$. Operation $\match$ is the most
basic one required to support almost every operation on parentheses-based tree
representations. 

Another basic query on parentheses is to compute the excess at some position
of $P$. We can reduce $\rank$ to it, because $\excess(B,i) = 
\rank_1(B,i)-\rank_0(B,i) = 
2\cdot \rank_1(B,i)-i$, thus $\rank_1(B,i) = (\excess(B,i)+i)/2$.

Because bit updates and queries reduce to a constant number of parentheses
updates and queries, we have the following result.

\medskip
\begin{corollary} 
Consider the problem of maintaining a sequence of balanced parentheses under
the operations of inserting and deleting a matching pair, and either queries
$\match$ or $\excess$.
In the cell probe model, with computer words of size $w=\log n$, \new{some of
those operations} on $n$ parentheses, where the fraction of updates is $1/q$, 
\new{must require} $\Omega(\log(n/q)/\log\log n)$ \new{time in the worst case.}
\end{corollary}

\section{Conclusions and Future Work}

We have shown how to store a dynamic bitvector $B[1\dd n]$ within (the
asymptotically optimal) $n+o(n)$ bits of space so that updates and queries 
can be solved in $O(\log(n/q)/\log\log n)$
amortized time if queries are $q$ times more frequent than updates. 
We have discussed applications of our result to a number of
dynamic compact data structures.

\new{We have proved that the above time is optimal {\em in the worst case},
so our {\em amortized} time is not optimal. For example, if we could know $q$ 
in advance (which our structure does not need) we could use the structure of 
Hon et al.~\cite{HSS03} with parameter $b=q$. Queries then need $O(\log_b n) 
= O(\log_q n)$ time, and updates need time $O(q)$, but since their relative 
frequency is $1/q$, their amortized time is constant.}

A first challenge for future work is to deamortize the times of our data
structure. This \new{would make it optimal, yet simultaneously retaining 
asymptotically optimal space} is not easy because in most cases deamortization 
implies building a new version of a structure while the old one is still in 
use, so both coexist and add up their space usage.

Another idea for future work \cite{JR12} is to analyze the use of 
splay trees instead of our weight-balanced trees, which should favor cases where
certain areas of the bitvector are frequently accessed. Our amortized analysis 
should then be combined with that of splay trees in order 
to ensure $O(\log(n/q))$ amortized time, while at the same time enjoying some 
of the (proven or conjectured) properties of splay trees \cite{ST85}. We note 
that such properties would not 
hold for individual bitvector positions, as they are packed in leaves of up to 
$b$ elements, but they could hold for sufficiently coarse bitvector areas.

Finally, in terms of functionality, we have not considered the problem of
cutting and concatenating bitvectors. This has been solved in time
$O(\log^{1+\epsilon} n)$ for any constant $\epsilon>0$ \cite{NS14}. It is easy
to implement them via flattening, but obtaining (poly)logarithmic amortized 
times that are also adaptive to $q$ is a challenge for future work.

\bibliography{paper}

\end{document}